\documentclass[runningheads,a4paper]{llncs}

\usepackage{amssymb}
\usepackage{graphicx}
\usepackage{subfigure}
\usepackage[usenames,dvipsnames]{xcolor}
\usepackage{verbatim}
\usepackage{algorithm}
\usepackage{algpseudocode}
\usepackage{multirow}
\usepackage{alltt}
\usepackage{array}

\usepackage{listings}
\usepackage{epstopdf}
\usepackage{epsfig}

\usepackage[hyphens]{url}
\urldef{\mailsa}\path|{aalsum,mweigle,mln}@cs.odu.edu|
\urldef{\mailsb}\path|herbertv@lanl.gov|

\newcommand{\keywords}[1]{\par\addvspace\baselineskip
\noindent\keywordname\enspace\ignorespaces#1}

\begin{document}

\mainmatter  

\title{Profiling Web Archive Coverage for Top-Level Domain and Content Language}

\titlerunning{Profiling Web Archive Coverage}

%
%
\author{Ahmed AlSum$^{1}$ \and Michele C. Weigle$^{1}$ 
\and Michael L. Nelson$^{1}$  \and \\ Herbert Van de Sompel$^{2}$}
%

\institute{$^{1}$Computer Science Department,
Old Dominion University,
Norfolk, VA USA\\
\mailsa\\
$^{2}$Los Alamos National Laboratory,
Los Alamos, NM USA\\
\mailsb\\
}
\authorrunning{A. AlSum et al.}

%
%

\toctitle{Lecture Notes in Computer Science}
\tocauthor{Authors' Instructions}
\maketitle

\begin{abstract}

%
%

 The Memento aggregator currently polls every known public web archive when serving a request for an archived web page, even though some web archives focus on only specific domains and ignore the others.  Similar to query routing in distributed search, we investigate the impact on aggregated Memento TimeMaps (lists of when and where a web page was archived) by only sending queries to archives likely to hold the archived page.  We profile twelve public web archives using data from a variety of sources (the web, archives' access logs, and full-text queries to archives) and discover that only sending queries to the top three web archives (i.e., a 75\% reduction in the number of
queries) for any request produces the full TimeMaps on 84\% of the cases.

\end{abstract}

\section{Introduction}
\vspace{-1em}

The web archive life cycle started with crawling the live web, then preserving for future access \cite{Brown2006}.
The global archived web corpus  is distributed between various web archives around the world. Every archive has its own initiative to crawl and preserve the web \cite{Shiozaki2009}, and these rules control its selection policy to determine the set of URIs for the web archive to crawl and preserve \cite{Masanes2006}. 


However, neither  the selection policy  nor the crawling log may be publicly available. This means that there is no way to determine what has been planned nor actually  archived.
This challenges our ability to search for a URI in the archives.  
  For example, the British Library Web Archive is interested in preserving UK websites (domains ending with .uk or websites existing in the UK)\footnote{\url{http://www.bl.uk/aboutus/stratpolprog/coldevpol/index.html}},  so searching it for The Japan Times\footnote{\url{http://www.japantimes.co.jp/}} may not return anything because that URI is not in the BL's selection policy.
  Furthermore, although  \url{www.bbc.co.uk} is covered in the BL web archive, a request for this page from the year 2000 should not be sent to the BL because it did not begin archiving until 2007.


For each web archive, we can determine a set of characteristics that distinguish the archive from  other archives and provide an insight about the archive content, e.g., the age of the archived copies and the supported domains for crawling. This profile enables the user to select the archives that may have the required URI at the specific ``datetime''. The application of this could be that the user-agent or the web archive could redirect the request based on the requested URI characteristics to another web archive that may have the URI. Also, the profile may help to determine the missing portion of the web that needs more coverage. 
 


Ainsworth et al.  \cite{Ainsworth2011} showed that between 16\% - 79\% of the web has been archived. The experiment was conducted between 13 archives and search engine caches. The results showed that the maximum number of archives that responded to the same URI was only 10 archives, so there was no single URI that appeared in all of the archives.

In this paper, we performed a quantitative study to create profiles for  12 web archives around the world (see Table \ref{tab:webarchivelist}). 
To build these profiles, we use a dataset constructed from URIs for the live web, fulltext search of the archives themselves, and access logs of the archives. We evaluated the constructed profiles in query routing between the various archives.

The rest of the paper is organized as follows. Section 2 describes the related work. Section 3 defines the archive profile characteristics. Section 4 defines the URI samples. Section 5 describes the experiment set and the results. Section 6 evaluates the usage of the profile in query routing. Section 7 concludes with a summary and future work for this study.

\begin{table}[tb]

\centering
\caption{List of Web archives under experiment.}
\begin{tabular}{ll  >{\centering\arraybackslash}p{1.5cm} l}\hline
&\textbf{Archive Name} & \textbf{FullText search}  & \textbf{Website}\\ \hline

IA & Internet Archive	& 	& \url{web.archive.org} \\
LoC & Library of Congress					& 	& \url{www.loc.gov/lcwa} \\
IC&Icelandic Web Archive &  & \url{vefsafn.is}\\ 

CAN & Library \& Archives Canada & x & \url{www.collectionscanada.gc.ca}\\
BL & British Library& x & \url{www.webarchive.org.uk/ukwa}\\
UK & UK Gov. Web Archive 		& x & \url{webarchive.nationalarchives.gov.uk}\\
PO&Portuguese Web Archive & x & \url{arquivo.pt}\\
CAT &Web Archive of Catalonia	& x & \url{www.padi.cat}\\
CR & Croatian Web Archive  & x & \url{haw.nsk.hr}\\
CZ&Archive of the Czech Web & x & \url{webarchiv.cz}\\
TW&National Taiwan University & x & \url{webarchive.lib.ntu.edu.tw}\\
AIT& Archive-It & x& \url{www.archive-it.org} \\\hline

\end{tabular}
\label{tab:webarchivelist}
\end{table}

\vspace{-1em}
\section{Related Work}
\vspace{-1em}

The general web archiving procedures have been studied by Masan\'{e}s \cite{Masanes2006}, Brown \cite{Brown2006}, and Br\"{u}ger \cite{NielsBrugger2005}.
Evaluating the current status of  web archives has been studied in various research. 
Shiozaki and Eisenschitz \cite{Shiozaki2009}
published a questionnaire survey conducted between 16  national libraries to justify the web archiving activities in the national libraries.
Niu \cite{Niu2012,Niu2012a} evaluated several web archives to study the  selection, acquire, and access techniques of the web archives. Niu limited her study to  web archives with an English interface. 

National libraries have published their  web archiving initiatives in various studies, for example, 
National Library of France \cite{Aubry2010}, 
Portuguese web archive \cite{Gomes2008}, 
National Library of the Czech Republic \cite{Vlcek2008}, 
National Taiwan University \cite{Chen2008},  
National Archives of Australia \cite{Heslop2002}, 
and China Web InfoMall \cite{Yan2004}.



Memento  \cite{VandeSompel2011} is an extension for the HTTP protocol to allow the user to browse the past web as the current web. 
The memento ($URI-M$)  is a snapshot for the original resource ($URI-R$) as it appeared in the past and was preserved by a web archive. 
The time that the memento was observed (or captured) by an archive is known as \textit{Memento-Datetime}. The TimeMap is a resource from which
a list of URIs of mementos of the original resource is available. A Memento Aggregator \cite{Sanderson2010a} provides a single TimeMap for multiple archives. The Memento Aggregator depends on various proxies \cite{Sanderson2010}  that provide Memento support for third-party servers and non-memento compliant web archives.


\vspace{-1em}
\section{Archive Profile}\label{sec:ArchiveCharacteristics}
\vspace{-1em}

An archive profile is  a set of characteristics that describe the content of the web archive. The goal of this description is to give a high-level overview about the web archive. This overview will help the user, other archives, or third party services to select the best web archive in case  selection between different archives is required. 
Examples of these rules include the following:

\begin{itemize}
	\item \textbf{Age}:  describes the age of the holding of the web archive. It is defined by the Memento-Datetime of the oldest memento in the archive. It may differ from the web archive starting date. For example, Portuguese Web Archive project started in 2007 but they included preserved materials that captured before 1995\footnote{\url{http://sobre.arquivo.pt/how-to-participate/supplying-historical-portuguese-web-contents}}.
	
	
	\item \textbf{Top-level domain (TLD)}:  describes the supported hostnames and top-level domains by the web archive. Some web archives have a special focus  that will consider specific domains only. For example, Library and Archives Canada have  focused on the \texttt{.gc.ca} TLD.
	
	
	\item \textbf{Language}:  describes the supported languages by the web archive. It varies depending on the motivation of the web archive creation. The Internet Archive has a wide range of languages, while the Icelandic web archive focuses on content in the Icelandic language. 

	
	
	\item \textbf{Growth rate}: describes the growth of the web archive corpus in the number of original URIs and mementos through time.

\end{itemize}

\section{URI Dataset Samples}
\vspace{-1em}


We prepared various URI  sample sets to profile the web archives. We sampled URIs from three sources: live web, archive holding, and archive access logs.

Open Directory (DMOZ)\footnote{\url{http://www.dmoz.org}} is used for URIs on the live web. Recording web archives' fulltext search responses  represent what the web archives  have already acquired. Finally, sampling from  user requests  to the Internet Archive  and Memento Aggregator represents what the users are looking for in the past. 
In all the samples, we used the hostname to create a top-level URI. For example, \url{http://example.org/a/b.html} will be \url{example.org}. Each sample has   unique   hostnames, however the different samples may have an overlap of  hostnames.
\vspace{-1em}

\subsection{Sampling from the Web}
\vspace{-0.5em}

DMOZ  is an open source web directory that is  built by user submissions of URIs. We selected DMOZ because it is well-represented in web archives \cite{Ainsworth2011}. We created three samples from DMOZ data:
\begin{itemize}

\item \textbf{DMOZ Random sample}:
We randomly sampled 10,000 URIs from the total directory of more than 5M URIs.

\item \textbf{DMOZ controlled (TLD)}:
We classified the DMOZ directory's URIs by the TLD. 
For each TLD, we randomly selected  2\% of the available hostnames or 100 hosts whichever is  greater.
We limited the study to a specific set of TLDs that are distributed around the world. 
The total number of URIs in this sample was 53,526 URIs.

\item \textbf{DMOZ controlled (Language)}:
DMOZ provides a specific list of URIs per language. We extracted these URIs and selected randomly 100 URIs from each language. The study focused on a limited set of languages that represent  the world. The total number of URIs in this sample was 2,300 URIs.

\end{itemize}
\subsection{Sampling from the Web Archive}
\vspace{-0.5em}

Most of the web archives  provide fulltext search   in addition to URI-lookup or collection browsing. 
We used the fulltext search to discover the hidden content of the web archives by submitting various queries and recording the responses.

This sample aims to calculate the overlap between the different archives and avoid  biasing for  archives that use DMOZ as a URI source (such as the Internet Archive). In order to reach a representative sample, we used two sets of queries:
\begin{itemize}

\item \textbf{Top 1-gram}:
The first set of queries terms was extracted from Bing Top 100k words as they appeared in April 2010\footnote{\url{http://web-ngram.research.microsoft.com/info/BingBodyApr10_Top100KWords.zip}}. We randomly sampled 1000 terms where most of them were in English. 

\item \textbf{Top Query Languages}:
The second set of queries was taken from Yahoo! Search query logs for nine languages\footnote{\url{http://webscope.sandbox.yahoo.com/catalog.php?datatype=l}}. This dataset has the 1000 most frequent web search queries issued to Yahoo Search in nine different languages.  
As the query terms are not limited to the search engine languages, they may have other languages especially English (e.g., Apple was one of the top query terms in Japanese). We filtered each file manually to include the designated language only and exclude the common terms (e.g., Obama, Facebook). %

\end{itemize}

We issued each query to all web archives that support fulltext search (see table \ref{tab:webarchivelist})\footnote{UK Gov. Web Archive  has a problem in searching with unicode characters}, then we recorded the top 10 results, and filtered by the hostname only. 
Table \ref{tab:queryterms} shows the total number of unique hosts returned by querying each query set from the archive. The total column has the total number of unique hosts that were retrieved  by each archive. The total column provides an indication of the size of the web archive. 

%
%

\begin{table}[tb]
\centering
\caption{Total number of unique hostnames returned from the query terms.}
\begin{tabular}{l r r r r r r r r  r |r|| r } \hline
 & \multicolumn{10}{c||}{Top Query Languages search}& \multirow{2}{1.1cm}{ Top 1-Gram} \\
  & chi  & eng  & fre  & ger  & ita  & jpn  & kor  &  por  & spa & Total & \\ \hline
AIT  &  26  &  2066  &  3512  &  3837  &  3321  &  119  &  2  &  2434  &  2141  &  12617  & 3953\\
BL &  163  &  2354  &  2350  &  2240  &  2068  &  225  &  131  &  1940  &  2056  &  6430  & 3187 \\
CAN   &   49  &  800  &  804  &  646  &  601  &  77  &  113  &  580  &  514  &  1351  & 1107 \\
CR   &   54  &  706  &  697  &  703  &  701  &  74  &  19  &  599  &  600  &  1599 &  1201 \\
CZ   & 363  &  1782  &  1578  &  1695  &  1519  &  577  &  114  &  1310  &  1278  &  6081  &  3360 \\
CAT   &   28  &  2775  &  2496  &  2448  &  2280  &  209  &  129  &  2164  &  2429  &  8996 & 4241 \\
PO   &   91  &  2460  &  3603  &  3081  &  3113  &  53  &  69  &  3267  &  3177  &  14126 & 5004 \\
TW   &  357  &  178  &  176  &  165  &  157  &  106  &  7  &  198  &  119  &  1004  & 354 \\
UK  &  0  &  2698  &  2009  &  2049  &  2046  &  0  &   0 &  1903  &  1871  &  8261  & 3431 \\
 \hline

\end{tabular}
\label{tab:queryterms}

\end{table}

\subsection{Sampling from Users' Requests}
\vspace{-0.5em}

The third sample came from   users' requests to the past web as recorded by the log files. 

\begin{itemize}

\item \textbf{IA Wayback Machine log files}:
IA Wayback Machine (WM) \cite{tofel2007} is the access interface for the Internet Archive, and has 240B+ URIs\footnote{\url{http://blog.archive.org/2013/01/09/updated-wayback/}}. WM receives more than 90M+ hits per day \cite{Alnoamany}. 
We selected log files for one week from (Feb 22, 2012 to Feb 26, 2012). We used only the requests to mementos or TimeMaps. For each memento or TimeMap, we extracted the original resource. We then sampled 1,000 URIs randomly from this list.

\item \textbf{Memento Aggregator logs}:
we sampled 100 unique hosts from the LANL Memento aggregator\footnote{\url{http://mementoproxy.lanl.gov/}} logs between 2011 to 2013.
\end{itemize}

\begin{figure}[hbtp]
\centering
\includegraphics[width=5in]{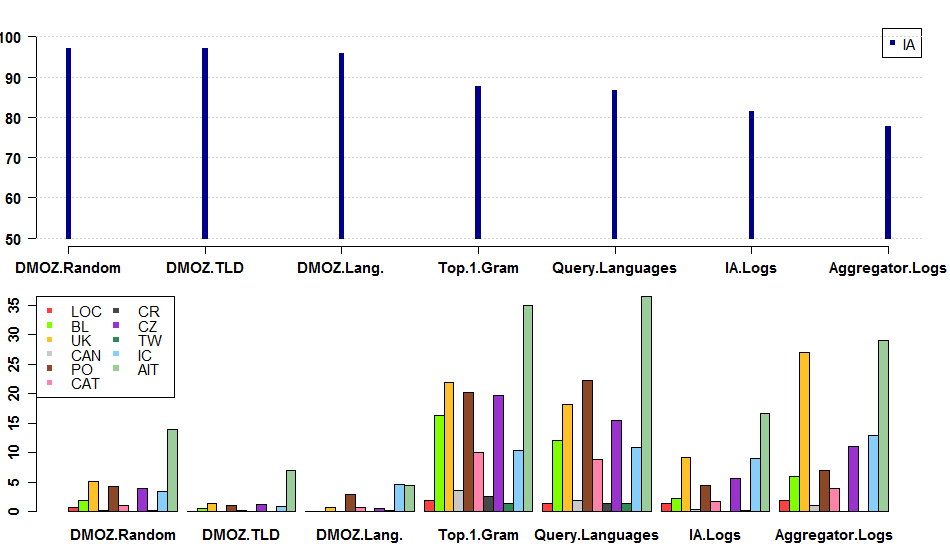}
\caption{Coverage histogram for all samples, IA on the top and all other archives below.}
\label{fig:all_archive_hist}
\end{figure}

\vspace{-2em}
\section{Experiment and Results}
\vspace{-1em}

For each hostname in the sample set (e.g., example.org), we converted it into a URI (i.e., http://example.org). We used the Memento proxies to retrieve the TimeMap for each URI. We recorded each memento with its Memento-Datetime. 
The coverage is the percentage of  URIs that were found in each archive  related to the sample size. In a future study, we will include other measurements such as the number of mementos per URI, the total number of mementos per archive, and the archive density \cite{Ainsworth2011}.

\textbf{Coverage}: 
Figure \ref{fig:all_archive_hist} shows the coverage for each sample through the archives.   The results show  IA, shown in the top graph, has the best coverage for all samples,  ranging from 79\% to 98\%. IA covered DMOZ samples with more than 95\% because DMOZ is used as a URI source for IA. The bottom graph shows the coverage for the rest of the archives. UK, PO, CZ, and IC show good coverage for specialized archives.

\begin{table}[tb]
\centering
\caption{The coverage percentage across the archives for fulltext search samples.}

\begin{tabular}{c| r r r r r r r r r r r r}\hline
 & \multicolumn{12}{c}{Target Archive} \\ 
Source  & IA & AIT & BL & CAN & CR & CZ & CAT & PO & TW & UK   & IC & LoC \\ \hline
AIT &	89.48	&	 \textbf{84.83} 	&	5.23	&	0.27	&	0.01	&	10.49	&	4.42	&	12.6	&	0.47	&	 19.17 	&	12.88	&	2.29\\ 
BL 	&	93.61	&	35.13	&	 \textbf{76.78} 	&	0.26	&	0	&	10.25	&	3.57	&	12.57	&	0.49	&	 40.99 	&	11.96	&	1.46\\ 
CA 	&	84.13	&	26.12	&	0.94	&	 \textbf{78.93} 	&	0	&	3.91	&	0.12	&	1.59	&	0.04	&	 7.24 	&	1.71	&	0.57\\ 
CR 	&	96.36	&	11.96	&	2.54	&	0	&	\textbf{ 52.93} 	&	4.61	&	1.68	&	3.71	&	0.29	&	 5.96 	&	4.00	&	0.32\\ 
CZ 	&	91.26	&	13.66	&	3.27	&	0.23	&	0	&	 \textbf{82.95} 	&	2.01	&	6.84	&	0.34	&	 7.50 	&	6.94	&	1.23\\ 
PA 	&	80.58	&	21.50	&	3.42	&	0.11	&	0.02	&	5.53	&	 \textbf{39.47} 	&	8.74	&	0.11	&	 10.83 	&	7.60	&	1.22\\ 
PO 	&	82.45	&	21.20	&	3.69	&	0.08	&	0.02	&	7.10	&	3.55	&	\textbf{ 58.94} 	&	0.14	&	 11.14 	&	10.23	&	1.29\\ 
TW 	&	93.00 	&	18.92	&	2.72	&	0.52	&	0	&	5.08	&	0.88	&	4.79	&	 \textbf{67.89} 	&	 8.17 	&	3.90	&	1.40\\ 
UK 	&	81.87	&	35.82	&	14.13	&	0.27	&	0.05	&	12.68	&	6.09	&	17.92	&	0.47	&	 \textbf{40.85} 	&	18.07	&	2.34 \\ \hline 
 
\end{tabular}
\label{tab:fulltextsample}
\end{table}

\textbf{Cross Coverage}: 
Figure \ref{fig:all_archive_hist} also shows that the web archives have a good coverage for the Top 1-Gram and Query Languages samples. This is because each archive contributed a part of the URIs to the samples as shown in table \ref{tab:queryterms}. 
Table \ref{tab:fulltextsample} shows the details of the coverage across archives. The table lists the URIs' source archives (as appeared in Table \ref{tab:queryterms}) on the rows and the queried archive on the columns. We can conclude the following from the table: 
\begin{enumerate}

\item The overlap between the web archives and IA is  high, which means that IA is effectively  a superset for all the archives. 
\item The overlap between the archives and each other is  low, which means they are covering different portions of the web. The highest overlap was between BL and UK because both   are focusing on UK domains.
\item

The web archives may have inconsistent interfaces between fulltext search and URI-lookup, as the overlap between the archive and itself is less than 100\% (highlighted in bold in table \ref{tab:fulltextsample}).
For example, querying BL with URIs that have been extracted from the BL fulltext search returned only 75\%. 
One reason   may be that we removed the URI path from the extracted URI, i.e., if fulltext search returned  (www.example.com/a/b.html), we will use (www.example.com) in the sample set.
 For the collection based archive, the curator may be interested in a specific URI and not the hostname itself. 
 
For example, 
you can extract \url{http://www.icsc.org.uk/index.html} from BL by searching for the term ``\texttt{Consumer Sciences}'', but there are no mementos for \url{http://www.icsc.org.uk/}. 
Another reason is that the web archive may index  the entire crawled corpus and make it available through  fulltext search (including the embedded resources). These embedded URIs may not be available through the URI-lookup interface. For example, Croatian Web Archive responds with embedded videos from \url{youtube.com} that  can not be discovered using the URI-lookup.

\end{enumerate}

\begin{figure}[tbhp]
\centering
\includegraphics[width=5in]{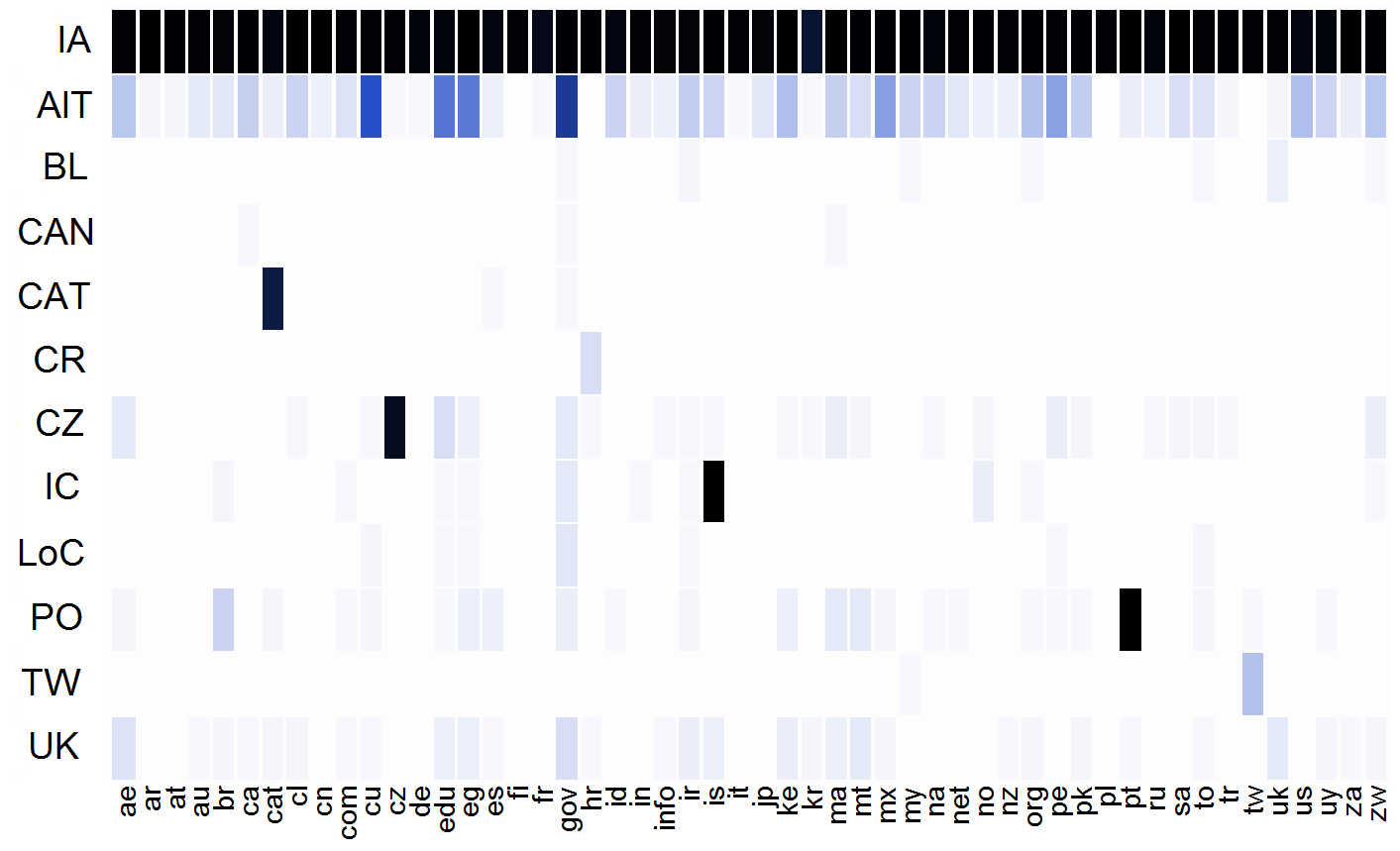}
\caption{Heat map of archive coverage for TLD samples.}
\label{fig:tld_heatmap}
\end{figure}

\begin{figure}[htbp]
\label{fig:tld_barplot}
	\centering
		\subfigure[The distribution of TLD per archive (Fulltext search).]{
		\includegraphics[width=1.0\textwidth]{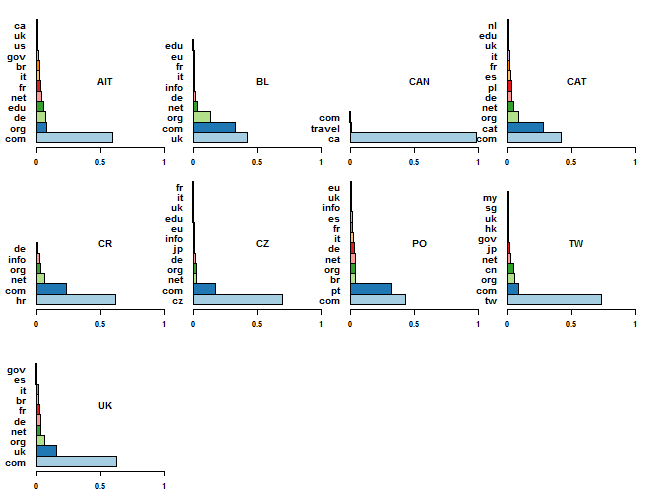}
	\label{fig:tld_coverage_search2}
	}
	\subfigure[The distribution of TLD per archive (DMOZ TLD sample). ]{
	\includegraphics[width=1.0\textwidth]{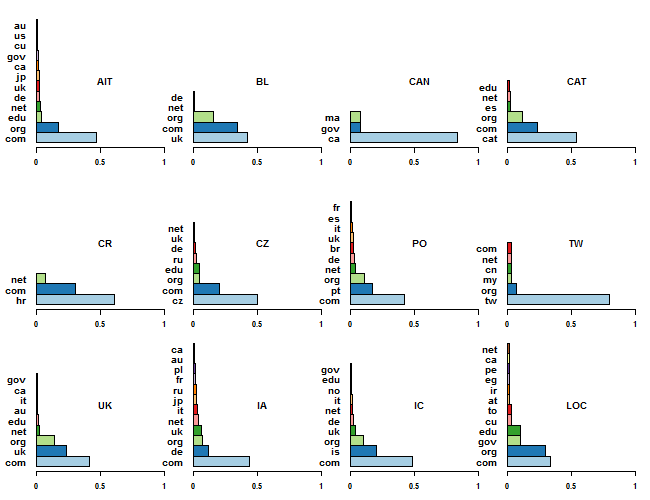}
	\label{fig:tld_coverage_sample}
	}
\caption{The distribution of the TLD through the archives.}
\end{figure}

\begin{figure}[htbp]

\centering
\includegraphics[width=1.0\textwidth]{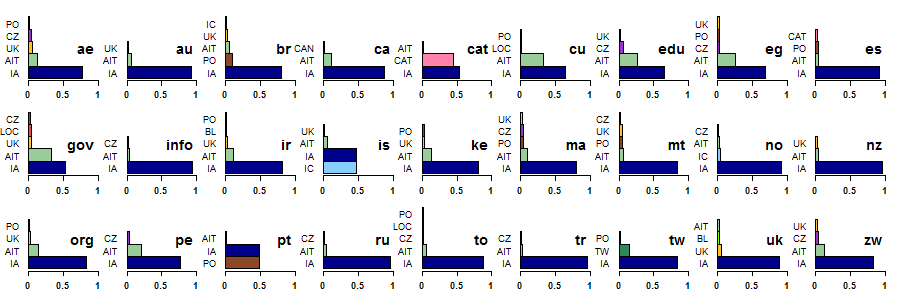}
\caption{Top-level domains distribution across the archives.}
\label{fig:tld_top_archives}
\end{figure}

\textbf{Top-Level domain distribution}:
Figure \ref{fig:tld_heatmap} shows the coverage of each TLD sample (columns) by the web archives (rows). 
``\textit{White}'' means the archive has 0\% of this TLD, ``\textit{Black}'' means the archives returned 100\% of the sample TLD.
The results show that IC, PO, CAT, TW, CZ, CR have good coverage for their national domains, and IC, PO, and UK extend their crawling beyond these domains. This behavior has been elaborated by studying the distribution of successfully retrieved URIs from each archive. 
Figure \ref{fig:tld_coverage_sample} shows the top TLD per archive from both  the fulltext search and the DMOZ TLD sample. There is a high correlation between both interfaces. The results show that even though the national archives work perfectly on their domains, they are not restricted to these domain only. CAN is the only closed archive for its domain. TW  supports a set of regional domains (i.e., .cn, .jp, and .sg).
Figure \ref{fig:tld_top_archives} illustrates the top archives per domain. It shows IA and AIT have high level coverage over the other archives for the general domains, however the national archives are doing similar or better  for their domains (e.g., CAT for \texttt{.cat} and IC for \texttt{.is}).

\begin{figure}[hbtp]
\centering
\includegraphics[width=0.8\textwidth]{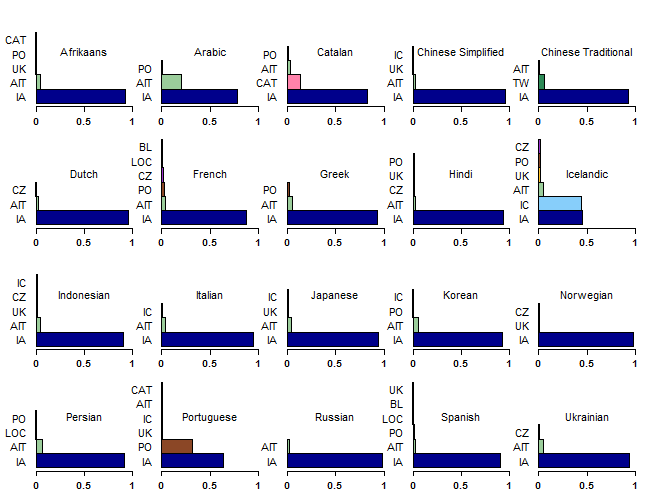}
\caption{Languages distribution per archive using DMOZ Language sample.}
\label{fig:lang_archive_hist}
\end{figure}


\textbf{Language distribution}:
Figure \ref{fig:lang_archive_hist} shows the coverage for each web archive divided by the language. CAT, IC, TW, and PT show  good coverage for their languages.

%
%
%
%

\begin{figure}[bhtp]

\centering
\includegraphics[width=1.0\textwidth]{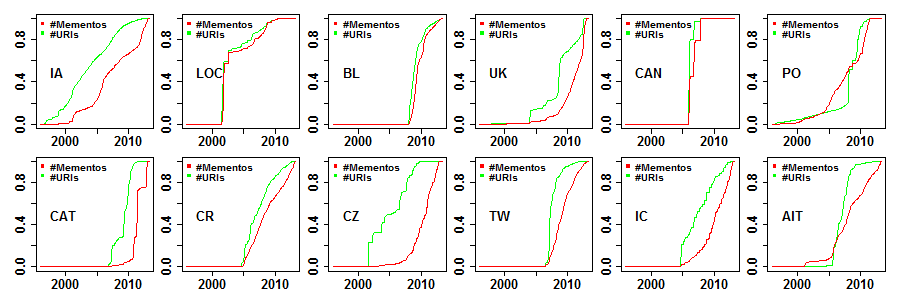}
\caption{Web Archive's corpus growth rate for URIs and Mementos.}
\label{fig:growthrate}
\end{figure}

\textbf{Growth Rate}: 
Figure \ref{fig:growthrate} shows the growth rate for each archive through  time. The growth rate is accumulated and normalized for the number of mementos and the number of new URIs added each month. The figure shows  both LOC and CAN covered a limited period of time, then they stopped their crawling activities as their number of new URIs does not increase. CZ and CAT stopped  adding new URIs a few years ago, but they are still crawling more mementos through  time. This figure also gives an idea about the start date for each archive. For example, IA and PO are the only archives that began before 2000.



%

\begin{figure}[bt]
\centering
\includegraphics[width=4in]{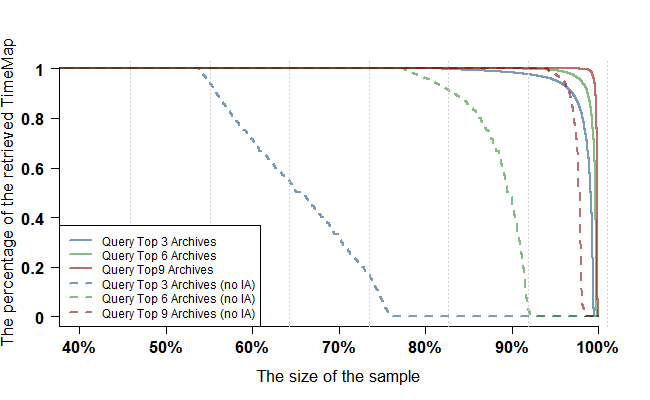}
\caption{Query routing evaluation using TLD profile.}
\label{fig:evaluation}
\end{figure}


Reflecting on the findings presented in this section, it is clear that IA is large, with mementos for 90\% of the dataset, and AIT and PO are in second place with 10\%.  This could be due in part to a bias in the dataset toward IA holdings, with logs from the Wayback Machine and the Mememento Aggregator as well as from DMOZ, a known seed URI site for IA.  Although we attempted to include content from a variety of archives, producing an unbiased dataset is difficult \cite{Ainsworth2011}.  Another possible explanation is simply that IA and PO have the oldest holdings, as both of them carried mementos from 1996.

There are surely other public web archives that exist that we simply did not know of.  Some regions of the world do not appear to have active, public web archiving projects  such as India and Africa. There are on going projects for the Arabic content by Bibliotheca Alexandrina and Latin America by the University of Texas\footnote{\url{http://lanic.utexas.edu/project/archives/}}. Finally, starting at about
2005 there appears to be a watershed moment for web archiving, when many projects begin to significantly grow their collections.



\vspace{-1em}
\section{Evaluation}
\vspace{-1em}

%
%
%
%
%
The profiles of web archives  could be used in optimizing  query routing for the Memento Aggregator. In order to quantify the success of the new profile, 
we applied ten-fold cross-validation.
We used TLD information (from Figure \ref{fig:tld_top_archives}) to create a general profile for the relationship between TLDs and archives. For each URI, we queried the aggregator with a different level of confidence using the top 3, top 6, and top 9 archives based on the requested URI's TLD.

We define the success criteria as how many URIs we get from a TimeMap when we select the top archives only.
For example, if using the top 3 archives retrieved 10 mementos and using the full TimeMap (all 12 archives) retrieved 15 mementos, we computed success as 0.67. 
Figure \ref{fig:evaluation} shows the normalized results for each case. 
We ran the experiment with all 12 archives and then repeated it IA.

The results show that we were able to retrieve the complete TimeMap in 84\% of the cases using only the top 3 archives. 
This increased to 91\% when using the top 6 archives.

Excluding IA, we were still able to get the complete TimeMap using the top 3 archives in 52\% of the cases. We assume each web archive query costs the same time for serving the request. In future work, we will profile the performance of responding to web archive queries. 

\vspace{-1em}
\section{Conclusions}
\vspace{-1em}

In this paper, we proposed an automatic technique to construct profiles for web archives. 
 The results showed that the Internet Archive is the largest and widest in  coverage. The national archives have good coverage of their domains and languages, and some of them extend their selection policies to cover more domains. The evaluation of using the profile in query routing retrieved the complete TimeMap in 84\% of the cases using only the top 3 archives.

In a future study, we plan to profile more characteristics such as: \textit{robots respect}, \textit{crawling frequency}, and \textit{crawling depth}. Also, we will use fulltext search to profile more characteristics in addition to the URI-lookup. In the evaluation, we will include more characteristics and increase the coverage of the sample URIs.

\bibliographystyle{splncs}
\bibliography{summary}
\end{document}